# PRECISION SPACE ASTROMETRY AS A TOOL TO FIND EARTH-LIKE EXOPLANETS

**A WHITE PAPER**

Michael Shao[1], Slava G. Turyshev[1], Eduardo Bendek[2], Debra Fischer[3], Olivier Guyon[4], Barbara McArthur[5], Matthew Muterspaugh[6], Chengxing Zhai[1], and Celine Boehm[7]

[1]*Jet Propulsion Laboratory, California Institute of Technology, 4800 Oak Grove Drive, Pasadena, CA 91109-0899, USA*

[2]*NASA Ames Research Center, Moffett Field, CA 94035, USA*

[3]*Yale University, New Haven, CT 06520, USA*

[4]*University of Arizona, Tucson, Arizona 85721, USA*

[5]*University of Texas at Austin, 2515 Speedway, Austin, Texas 78712 USA*

[6]*Tennessee State University, 3500 John A. Merritt Boulevard, Nashville, TN 37209, USA*

[7]*Durham University, Stockton Road, Durham, DH1 3LE, UK*

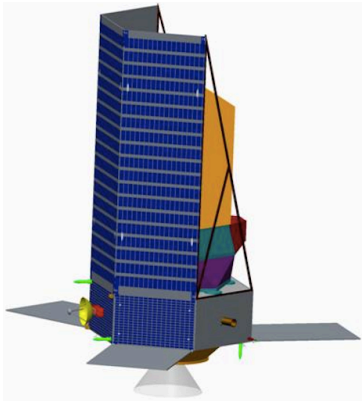

*Figure 1. MAP satellite concept.*

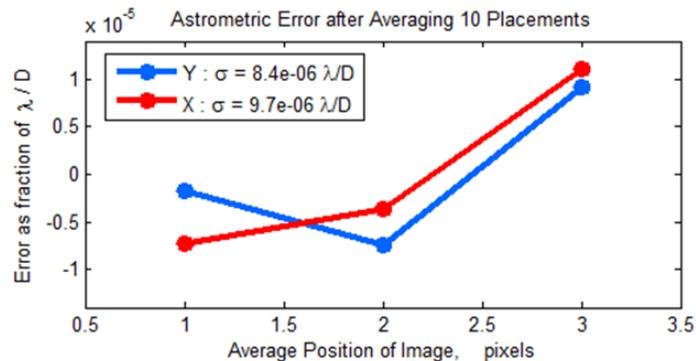

*Figure 2. Pixel centroiding has been demonstrated to ~1e-4 pixels in a single image and 3e-5 in an average of 10 images.*

The Microarcsecond Astrometry Probe (*MAP)* mission concept is designed to find $1M_\oplus$ planets at 1AU orbit (scaled to solar luminosity) around the nearest ~90 FGK stars. The *MAP* payload includes i) a single three-mirror anastigmatic telescope with a 1-m primary mirror and metrology subsystems, and ii) a camera. The camera focal plane consists of 42 detectors, providing a Nyquist sampled FOV of 0.4°. Its metrology subsystems ensure that *MAP* can achieve the 0.8 μas astrometric precision in 1 hr, which is required to detect Earth-like exoplanets in our stellar neighborhood.

Michael Shao, (818) 354-7834, michael.shao@jpl.nasa.gov

# 1 Introduction

The ultimate goal of exoplanetary science is to answer the enigmatic question "Are we alone?" via unambiguous detection of biogenic gases and molecules in the atmosphere of an Earth's twin around a Sun-like star. Directly addressing the age-old questions related to the uniqueness of the Earth as a habitat for complex biology is one unprecedented, cross-technique, interdisciplinary endeavor. What is most important is that our generation could answer this question in the near future.

As is known, there are several techniques exist to find exoplanets, with most frequently used are radial velocity (*RV*), and planetary transits. Even though these two techniques have succeeded in finding 1,000s of exoplanets (with most them being Neptune or super-Earth sized planets), the ultimate goal of finding an Earth's twin in a 1AU orbit around a solar-type star remains unfulfilled.

The goal of the NASA Kepler mission was to determine $\eta_\oplus$, the rate of occurrence for Earth-like planets, i.e., planets similar in size and orbital radius as the Earth. The transit of an Earth-like planet in front of the Sun causes the solar flux to dim by a factor of $10^{-4}$. Due to a combination of slightly increased instrumental noise and stellar photometric variability, Kepler needed about 8 years to see 8 transits needed to build up the signal-to-noise ratio (SNR) for detecting a $1R_\oplus$ planet transiting a G star. However, with the failure of 2 out of 4 reaction wheels that happened in ~4 years after launch, this goal was abandoned.

Since Kepler multiple transit missions were approved, e.g. *TESS*, *Cheops*, *Plato*. However, none of them plan to monitor a star field continuously for 8-yr mission lifetime, which is necessary to detect $1R_\oplus$ planets in a 1-yr orbit. These transit missions are all aimed at short period planets. Earth-sized planets in the HZ of M stars are certainly very interesting, but planets that are tidally locked to the parent star are very different from our Earth. If our Earth were orbiting an M star, it would likely be tidally-locked, so that the water would evaporate from its sunlit side and would condense as ice on the dark side. While our Earth is ¾ covered with water, that water is only a few miles deep. On a tidally-locked Earth, the dark side would have a few miles of ice, a very large Antarctica. Thus, it is hard to imagine that life could exist on such a tidally-locked planet.

Currently, the best accuracy achieved by *RV* technique is ~0.4 m/s. An Earth-Sun system would have a *RV* signal of 0.09 m/s. Multiple ultra-high precise instruments are under construction with the goal of achieving the accuracy of 0.1 m/s. But the astrophysical noise due to stellar activity such as radial pulsations could affect *RV* at ~0.3 m/s. Methods to model and compensate/correct the astrophysical noise for *RV* measurements current is an ongoing research topic. Also *RV* method can only determine the product $M^*sin(i)$, but not the planet's mass nor the orbital inclination *i*.

Measuring masses of exoplanets is important. Planets with masses $>5M_\oplus$ have enough gravity to hold on to a hydrogen atmosphere for billions of years. Measuring mass allows separating mini-Neptunes from super-Earths. Rocky planets in the solar system have a density much higher than that of gaseous planets. If a rocky exoplanet would have liquid water on its surface, it may also support life similar to that on the Earth. Thus, ability to measure the mass to $1M_\oplus$ without *sin(i)* ambiguity is important to find Earth's analogues.

While both *RV* and transit techniques are good at finding planets with very short orbits (period of days), astrometric measurement of stellar reflex motion of exoplanets, on the other hand, is more sensitive to longer orbits. For a terrestrial planet in the habitable zone (HZ) of nearby stars, a typical signal level is 0.3 μas. The possibility of an Earth transit is ~0.47%, making hard to detect. Thus, neither of these methods (*RV* or transit) is optimized for finding systems like our own.

Astrometry measures the reflex motion of the star in 2D, from which both the planet mass and orbit inclination can be derived. Unlike other methods, such as Doppler spectroscopy, astrometry is less sensitive to the disturbances due to stellar activity (spots, meso-structures). Astrometric detection of an exo-Earth at SNR=6 would result in a 1σ error on the planet mass of ~0.25 $M_\oplus$. As such, precision astrometry is a unique tool for discovery of exo-Earths.

Once we detect a planet we would want to detect the light from it and measure its spectra. Coronographic instruments on the next generation space telescopes (such as the concept studies for HabEx, LUVOIR, and OST) will obtain atmospheric spectra. However, the simulations for these studies show that independent detection will make these missions more efficient, saving precious 10-m class space telescope time for characterization. However, is it more economical to use astrometry to "discover" exo-Earth? We think that the answer to this second question is also yes.

How many stars one has to search before finding ~10 exo-Earth? While the Earth's mass and its radius are well defined quantities, the definition of the HZ is less clear. Event our own Earth would not have liquid water if it were in the Venus's or Mars' orbits. Since the fraction of stars in the HZ depends on the "width" of the zone, there is still a debate on the definition of HZ. Is there a value for astrometry mission "preceding" a coronagraphic mission? We argue that there is.

A concept like *LUVOIR* is so large that it would detect exo-Earths both astrometrically and via direct detection. But the diameter of a coronagraphic telescope is linearly related to the distance of that Earth 2.0. Knowing where the nearest half-dozen of exo-Earths are, lets one design a coronagraphic mission that can get spectra of those known exo-Earths. The cost of a space telescope goes as ~$D^{2.7}$, so knowing where the targets are can significantly reduce the uncertainty in the cost of the coronagraphic mission. With our detector metrology, the ~9-15m aperture *LUVOIR* could do both direct detection of Earth 2.0 as well astrometric detection. However, an astrometric detection of a planet around the nearest ~100 stars would only take ~100 hrs of integration time on *LUVOIR*.

Unlike the Doppler and transit methods, *astrometry* alone can determine reliably and precisely the true mass and three-dimensional orbital geometry of the exoplanet system, which are fundamental inputs to models of planetary evolution, biosignature identification, and habitability. With recent advances in astrometric detector calibration techniques (Zhai et al., 2011, Crouzier et al 2016), newly-developed flight metrology techniques (Shao et al 2011), availability of the highly-precise astrometric catalogues (e.g., Gaia 2018), and existence of various mission concepts (Malbet, 2016), it is time to consider optical astrometry as a standalone technique for a space mission.

As an example, we consider a concept for the Microarcsecond Astrometry Probe (*MAP*), which would be a ~1m telescope with astrometric accuracy at the μas-level that is achieved by measuring and correcting systematic errors in the telescope structure and its focal plane (Figure 1). This technology enables sub μas-level astrometric measurements in ~1 hr of integration on candidate stars. As such, it plays a significant role in the ESA's proposed M5 mission *Theia*, it was considered for WFIRST, and, is used in the studies for *LUVOIR*. *MAP* has the potential to 1) discover most of the potentially habitable planets around the nearest ~90 stars to the Sun, 2) directly measure their masses and system architectures, and 3) provide the most complete target list and vastly improve the efficiency of detecting potential habitats of complex exo-Life with the next generation space telescopes and Extremely Large Telescopes (ELT).

*Table 1. MAP summary.*

| | |
|---:|:---|
| Duration | 5 years |
| Instrument | 1 m telescope |
| Field of view | 0.4 degree |
| Orbit | HEO/GEO or L2 |
| FP Detector | 42 CCDs + metrology |
| Astrometry | 0.8 μas (1 hr integr) |
| Wavelength | 0.4-1.0μm |
| Cost | $500M |

## 2 Technical Description of MAP

The baseline *MAP* payload is deliberately simple: it includes a single three-mirror anastigmatic telescope, with metrology subsystems, and a camera. It is designed to be a ~1m telescope with a large focal plane, 0.4° field of view (FOV) with point-spread function (PSF) Nyquist-sampled and metrology systems that would calibrate instrumental errors in the focal plane and optics at the µas level (Table 1). The observatory would be located in HEO/GEO or L2 to provide thermal stability, and high efficiency operations within the 5 years of mission lifetime. Its metrology subsystems ensure that *MAP* achieves the precision 0.8 µas in 1 hr integration time.

***Science requirement***: To detect a planet, position of the parent star must be observed many times and its reflex motion due to orbiting Earth detected. Numerous numerical simulation has shown that using a periodogram to detect an orbit requires a SNR of ~6, for the false alarm rate to be < 1%. Here SNR is defined as the amplitude of the periodic reflex motion of the star divided by *err/sqrt(N)*, where *err* is the 1 hr astrometric error and *N* is the total number of hours observed during the mission. An Earth-Sun twin at 10 pc would have a reflex amplitude of 0.3 µas. A SNR of 6 would require a total of 256 hrs of observation. Assuming that these 256 hrs would be distributed over the 5-year mission, we require our error budget to be 0.8 µas over 1 hr integration.

To detect the reflex motion, the observational period must be longer than the period of the planet. To calculate the number of stars where *MAP* can detect the presence of an Earth twin, we used the Hipparcos catalog and collected all the FGK stars with 30 pc, which is ~400 stars. We then placed a $M_\oplus$ planet in mid-HZ (1AU scaled to the luminosity of the star), calculated the astrometric signature, and sorted the list from easiest to most difficult for astrometric detection. We then calculated the of observing time needed to detect a $1M_\oplus$ planet for each target. As a result, we will allocate ~35% of the mission ~15,300 hrs for finding Earth twins, there would be sufficient time to search 89 stars. Assuming $\eta_\oplus$ in the HZ is 10% *MAP* should find ~9 such planets. These planets would be the prime target for a future direct detection mission such as *HABEX, LUVOIR*.

We achieve the required 0.8 µas by calibrating systematic errors and averaging down the random errors mainly due to the photon noise. For a star at 10 pc, the astrophysical noise is ~0.08 µas. This noise is due to features on the surface of the star that are asymmetrical as seen by *MAP*. For a Sun-like star with a 30-day rotation period, we expect this noise to be correlated on time scales <7 days, the time it takes for a star spot to rotate from one side to another or behind the star. The astrophysical noise decreases as $1/(N_{obs})^{1/2}$, where $N_{obs}$ is the number of observations separated by > 7 days. At this level, if we devote enough observing time, planets $½M_\oplus$ could be detected.

***Instrument***: Taking into account requirement above, the following program is proposed. Observations of the best nearest 90 FGK stars will be obtained with number of visits $N_{vis}$ = 50 and integration time $t_{vis}$ ~ 1 hr, plus a 0.20 hr slew time between targets (~10 deg). The total duration for such a program is 0.6 yr (~5,400 hrs), or 20% of a 3 year observing mission time. This program will also be valuable for understanding planetary diversity, the architecture of planetary systems (2D information plus the Kepler's laws, result in a 3D knowledge) including the mutual inclination of the orbits, a piece of information that is often missing in studies of planetary systems.

The science payload is a ~1m diameter TMA telescope, with a ~0.4° FOV. At the focal are the sCMOS detectors. If the telescope and detectors were perfect, we should be able to make astrometric measurements limited only by noise which is $\lambda/(2*D*[photometric\ SNR])$. When measuring the position of bright nearby stars, the photon noise of the bright stars usually dominates over the photon noise of the reference stars. In this "bright star" scenario, it is essential to control systematic errors. In an ideal telescope, the spherical sky projected onto a flat tangent plane is perfectly relayed to the 2D focal plane detector. Imperfections in the telescope produce optical distortion.

There are sources of detector error: imperfect pixel geometry, nonlinearity in the detector and readout electronics, electronic crosstalk between readout channels, non-uniform QE within a pixel.

In *MAP* the optics should be diffraction limited, with < $\lambda/20$ wavefront error across the FOV. Also, the focal plane will be Nyquist sampled, that is > 2 pixels per $\lambda*f/\#$ in the focal plane. The diffraction limit of a 1m telescope is $(\lambda/D)\sim0.12$ arcsec. With the objective of reaching the accuracy of ~1 μas or better, this means centroiding the stellar image to $\sim10^{-5}$ of the diffraction limit. Fortunately, neither the optics nor the detectors need to be "perfect" at the $10^{-5}\,\lambda$ level, but it is necessary to "calibrate" these systematic errors so that their biases can be removed in data analysis. In all cases, the calibration of instrumental errors is made easier if they are stable in time. This lead to placing the observatory in a stable thermal environment, such as GEO or L2 orbits. The other spacecraft requirement that's important for μas astrometry is stable telescope pointing. Micro-arcsecond astrometry needs very high SNR observations, e.g. SNR~$10^5$ for 1 μas accuracy. This high SNR means the measurement will involve averaging 1,000's of images. The telescope pointing has to be stable enough so that during a single exposure (~0.1-10 sec) the telescope should not drift by more than ~1/20 of the diffraction limit, about 6 mas.

***Detector calibration*** is necessary for high precision astrometry because focal plane array detectors like CCDs have inter-pixel response variations at the level of a few parts of a thousand to one percent. The leading order inter-pixel response is the flat field response, which can be calibrated by illuminating the detector with a uniform light. The next to leading order of inter-pixel response variation is the deviation of the effective pixel locations from a regular grid. To calibrate irregular effective pixel locations, we will use a laser metrology technology that we developed at JPL in 2010, where a pair of laser metrology beams shine at the detector providing an illumination of spatial fringes. We demonstrated a precision of $10^{-4}$ pixel for differential centroiding by accounting for the irregular pixel location calibration using an E2V chip. Figure 2 shows the results of the experiment that demonstrated centroiding to $10^{-5}\,\lambda/D$ after averaging a set of 10 images when measuring the separation of stars on a CCD with 4 pix/$(\lambda/D)$ sampling (or $1.2\times10^{-4}$ pix in 1 image).

***Field distortion calibration***: Field distortion in the optical system can be very large compared to 1 μas. If the field distortion is stable, it can be calibrated using reference stars in a manner similar to what's been done on HST. How stable does the wavefront has to be? We performed numerical simulations of centroiding errors when random wavefront errors are introduced. Introducing wavefront errors of $\sim0.5\times10^{-3}\,\lambda$, we found centroiding offsets of $\sim10^{-5}\,\lambda/D$. The reason there is not a 1:1 correlation between wavefront and centroid error is because "even" wavefront aberrations like focus, astigmatism, spherical aberration do not affect centroids. Also, high-order wavefront errors are inefficient in coupling to the centroid position. Roughly speaking, while a $10^{-10}$ coronagraph needs the wavefront to be stable to few picometers, microarcsec astrometry can be achieved with wavefront stability that is ~30-50 times larger. A space astrometric telescope, while it needs to be much more stable than HST, does not have to be as stable as an exo-planet coronagraph.

One area that can use further study is the complexity of or the number of degrees of freedom needed to model distortion for microarcsec astrometry. HST astrometry has been successful at 100 μas level with just modeling 3rd order distortion terms. How many terms are needed to model distortion at 1 μas? The diffractive pupil concept studied by Guyon and Bendek is a way that can be explored in the lab on the flight telescope or a prototype before launch, even if the diffractive pupil is not needed in orbit because the telescope can be built with the near coronagraphic stability.

***Technology readiness***: Most of the technologies for *MAP* are mature. For example, a large number of 1-m telescopes have been flown in space. There are ~2-3 technology areas where flight qualification should be addressed. One is the on-board metrology system to measure the geometry of the FPA. All of the components, such as lasers, modulators, fibers have been flown in space, but the system was not flown. The second technology is the diffractive pupil, should we decide to use it as our primary means for field distortion calibration. Diffractive optics is a fully passive device and there should be no issues with respect to flight qualification. The main technology challenge is putting it on a large optic. A potential third technology is a laser metrology system to measure changes in the alignment of the telescope/focal plane. Again, all the components were flown in space, but the system is not yet flight qualified. All key technologies are already available as a result of SIM, JWST, and LISA projects, presenting a lower level of risk.

The ***error budget*** is 0.8 µas over 1 hr observation for narrow angle differential astrometry using a 1m telescope. To understand the error budget, it useful to introduce our basic measurement, which is obtained with 30 sec integration tracking stars in the field. The total 1 hr integration consists of 120 basic 30-sec measurements. The 120 basic 30-sec observations are carried out by dithering the locations of the stars on the detector relative to each other. This randomize the residual systematic calibration errors from field distortion or detector response inter-pixel variations, thus enables to average down errors approximately as ~ 1/sqrt(120) to reduce the error by a factor of 11.

We require that a basic 30-sec observation should have ~8.8 µas accuracy to achieve the 0.8 µas after integration of 1 hr. The differential astrometry error can be understood as errors due to target star position and reference star measurement. We assume the target star is bright (mag 7 is assumed), so its noise is estimated ~2.8 µas given by $\sigma = W/(2\sqrt{N_{ph}})$, where $W$ is the size of PSF($\lambda/D$) = 0.12" and the total number of photon $N_{ph}$ is ~$4.8 \times 10^8$ for 30 sec integrations.

For target star, the error is dominated by systematic error from field distortion and detector calibration with a budget of 7.2 µas broken into ~5.16 µas each for detector and field distortion calibrations. We have demonstrated $1 \times 10^{-4}$ pixel precision in centroiding. With a plate scale of 1pixel = 50 mas to critically sample the PSF at 2.4 pixels, this error budget is met for detector response calibration. The field distortion is assumed to be the same size of the systematic error due to pixel cal. The key is the stability. For a stable system, by observing bright cluster, the field distortion can be calibrated to µas with detector calibrated to µas level.

There are less photons ref stars. We estimate the random error from ref stars is ~3.9 µas. The systematic error from each ref star is the same as the target star ~7.2 µas. Because we have over 900 ref stars located at different pixels on the detector, we expect the systematic error can be averaged down to 0.7 µas. This shows how the total error budget of 4 µas for ref stars can be met.

## 3 Conclusion

Because of the many recent technological advances, the key technologies needed for a µas-level astrometry mission are now in hand. Given many years of NASA support for the Space Interferometry Mission, more recent grants from DARPA, and USAF to develop the detection calibration, precision metrology, etc., precision astrometry is ready for the primetime in space at a cost similar to the Kepler mission. Such a mission would search ~100 nearby stars equivalent to $1.0 M_\oplus$ planets with a 1.0-yr orbit around nearby FG late K stars. This could provide ~10 specific targets for a much larger coronagraphic mission that would measure its spectra.

We ask the NAS Committee on an Exoplanet Science Strategy to consider astrometric mission concepts capable of finding Earth-2.0. We also ask the Committee to recommend NASA to consider development of such missions relying on precision astrometry starting in the next decade, perhaps in collaboration with other national space agencies.